\documentclass[smallextended]{svjour3}       
\smartqed  
\usepackage{graphicx}
\usepackage{mathptmx}      
\usepackage{amssymb}
\newcommand{\EQ}{\begin{equation}}
\newcommand{\EN}{\end{equation}}
\newcommand{\EQA}{\begin{eqnarray}}
\newcommand{\ENA}{\end{eqnarray}}

\newcommand{\Eq}[1]{Eq.~(\ref{#1})}

\newcommand{\Sec}[1]{Sec.~\ref{#1}}

\newcommand{\Fig}[1]{Fig.~\ref{#1}}

\newcommand{\Tab}[1]{Table~\ref{#1}}

\newcommand{\mean}[1]{\overline #1}

{}
\newcommand{\yjour}[4]{, #2 {\bf #3}, #4 (#1).}

\newcommand{\ybook}[3]{, {\em #2}. #3 (#1).}
\journalname{Heat and Mass Transfer}
\begin{document}

\title{Modeling radiation in particle clouds: 
    On the importance of inter-particle radiation for pulverized solid fuel 
    combustion}
\author{Nils Erland L.~Haugen \and Reginald E. Mitchell
}
\institute{Nils Erland L.~Haugen \at 
SINTEF Energy Research, NO-7465 Trondheim, Norway
\email{nils.e.haugen@sintef.no}
\and
Reginald E. Mitchell \at
Department of Mechanical Engineering, Stanford University, 
Stanford, CA 94305, USA}

\date{Received: date / Accepted: date}

\maketitle

\begin{abstract}
The importance of inter-particle radiation for clusters of gray and diffuse
particles is investigated.
The radiative cooling of each individual particle is found to vary strongly
with its position in the cluster, and a ``mean'' radiative particle cooling 
term is proposed for single particle simulations of particle clusters or for
high detail simulation, like Direct Numerical Simulations of small 
sub-volumes of large clusters of particles. Radiative cooling
is shown to be important both for furnaces for coal gasification and 
coal combustion. Broadening the particle size distribution
is found to have just a minor effect on the radiative particle cooling. This is
particularly the case for large and dense particle clusters where there is
essentially no effect of size distribution broadening at all. For smaller and
more dilute particle clusters, the effect
of distribution broadening is clear but still not dominant. 
\keywords{combustion\and coal\and radiation\and simulation\and particle} 
\end{abstract}

\section{Introduction}
Many industrial processes, such as {\it e.g.} pulverized
coal or biomass combustors, fluidized bed reactors or entrained flow reactors
rely on reacting particles. In order to fully understand these systems, an 
understanding of the chemical reactions together with the heat transport 
to and from the particles is crucial. In most cases, convective and conductive
heat transfer between the particles and the gas must be considered. For high
temperatures, radiative heat transfer should also be taken into account. Here
one can think of both particle-fluid interactions, particle-wall interactions and
particle-to-particle interactions. 
In the work reported here, the importance of particle-to-particle radiation
is discussed. 

When performing CFD simulations, particle radiation is often included and 
found to be important \cite{azad1981,park1998,boutoub2007}. If, on the other
hand, one does not perform a full CFD simulation but is rather interested 
in solving single particle physics and chemistry
in high detail one often neglects, or partly neglects, radiation. In such cases 
radiation may not be considered at all, or if it is taken into
account,
only particle-wall radiation \cite{qiao12,xu12,mitchell07} or particle-fluid
radiation \cite{liu01} is considered.
The primary aim of this paper is 
to obtain a realistic description for the particle
radiation transfer, including both particle-to-wall and particle-to-particle
radiation, that can be used for high detail particle simulations.
The secondary aim is to investigate the effect of particle size
distribution broadening on radiative transfer.

In the current work, only geometric scattering is considered, and the
analysis is limited to the case where the particles radiate like graybodies
and the gaseous environment between particles is transparent to radiation. 
Considering only geometric scattering is valid since 
the particles have
large size parameters, {\it i.e.} $\xi=2\pi r_p/\lambda>5$ where $\lambda$ is the 
wavelength of the radiation and $r_p$ is the particle radius, 
such that Rayleigh and Mie scattering
can be omitted.

Consider a cloud of hot particles embedded in a radiatively transparent
gas and enclosed within a confinement. This could for example resemble
the situation in an entrained flow gasifier.
If the
radiative flux absorbed by a particle is $F_a$ and the flux absorbed by
a replacement blackbody particle having the same size and temperature
is $F_{bb}$, then an absorption efficiency factor for the particle can be
defined as $E_a = F_a/F_{bb}$, which is a measure of the efficiency of the
particle as an absorber compared to that of a blackbody.

A ray of radiation incident on a large particle will either be
absorbed or reflected by the particle surface.
Since the total cross section of a particle with radius $r_p$ is 
$A_p=\pi r_p^2$, the absorption cross section must be $A_a=E_aA_p$ 
given that a fraction $E_a$ of all the radiation incident on the particle
is absorbed. 
Since radiation is either absorbed or reflected the scattering cross section
of the particle must be $A_s=A_p-A_a=(1-E_a)A_p$. 
A scattering efficiency factor is defined, analogously to the absorption 
efficiency factor, as the fraction of incident radiation that is scattered by
the particle surface
$E_s=A_s/A_p$, which then yields $E_s+E_a=1$. For the large particles
of interest, the scattering efficiency factor 
equals the reflectivity
of the particle surface while the absorption efficiency factor equals the
absorptivity of the particle surface. In all of the following the 
scattering efficiency factor of the particles 
is assumed to be much smaller than the absorption efficiency factor such 
that the effect of scattered radiation from the particles can be 
neglected. 

Performing three dimensional  CFD simulations of full gasifiers or combustors
are very demanding. Due to the large CPU power required one often has to use very 
simplified chemical models, both for the homogeneous and heterogeneous 
reactions. In many situations it is therefore better to simulate one
single particle with high fidelity chemistry, and let this particle 
represent the ``average'' particle in
the domain. With this simulation method one can easily do a large 
parameter scan over a range of different parameters with detailed chemical 
reactions. Such an ``average particle'' simulation will not yield detailed 
information of geometrical features in any application. Instead it will yield
qualitative trends, using accurate chemical kinetics, 
for a range of parameters in ``typical'' conditions 
relevant for the application of interest.
Traditionally, the particle cooling term used for such single particle 
simulations of a cloud of particles has been given 
by~\cite{qiao12,mitchell07}
\EQ
\label{eq1}
Q=A_p(q_p-E_aq_w)
\EN
where $q_p$ and $q_w$ are the thermal radiation from the particle and 
the wall, respectively. It is evident from this that inter-particle
radiation is neglected, which may not be a good assumption for many applications. 
A description of a particle cooling term
that does include inter-particle radiation for this kind of simulation tool 
does not exist in the open literature.
The main objective of the current work is therefore to extend the
above radiative cooling term to also take into account inter-particle radiation.

\section{The extinction coefficient for a cloud of particles}
\label{sec2}
The extinction coefficient is a measure of how easily a ray of radiation
penetrates a given medium without being absorbed. 
Let a large number of small particles be 
embedded in the fluid such that the number density of the 
particles with radius between $r_p$ and $r_p+dr_p$ is $n(r_p) \; dr_p$. 
If the particles are treated as diffuse 
graybodies with zero scattering coefficients, a
ray of radiation emitted from the source at $r=0$ may be absorbed by 
the particles.
The probability of extinction depends on the number density 
of particles, the projected particle surface area and the length of travel.
The extinction coefficient, $K$, of the medium due to the embedded particles is 
given by 
\EQ
\label{abs_coeff}
K=\int_{r_p=0}^\infty (E_a+E_s) n(r_p) \pi r_p^2 dr_p
=\int_{r_p=0}^\infty n(r_p) \pi r_p^2 dr_p.
\EN 

Let's now assume a Gaussian particle size distribution given by
\EQ
\label{eq2}
n(r_p)=\frac{n_p}{\sigma_p\sqrt{\pi}}\exp\left(-\left(\frac{r_p-\bar{r}_p}{\sigma_p}\right)^2\right),
\EN
where $n_p$ is the total particle number density, $\bar{r}_p$ is the mean
particle radius and $\sigma_p$ is the width of the particle size distribution.
It is convenient to define the distribution 
width as a fraction $\gamma_p$ of the mean
particle radius $\bar{r}_p$, {\it i.e.} $\sigma_p=\bar{r}_p \gamma_p$. 
Employing this in \Eq{eq2}, and using the result in \Eq{abs_coeff} yields the
following expression for the extinction coefficient
\EQ
\label{ext_coeff}
K=\pi n_p\bar{r}_p^2
\left[
1+\frac{\gamma_p}{\sqrt{\pi}}+\frac{\gamma_p^2}{2}
\right].
\EN

The equation of radiative transfer, which describes the change in 
spectral radiative intensity with $s$ around the wavelength $\lambda$ 
in the solid angle $d\omega_i$ about the 
direction of $s$, is given by
\EQA
\frac{dI_\lambda(\lambda,s)}{ds}&=&
-a_\lambda I_\lambda(\lambda,s)
+a_\lambda I_{\lambda,b}(\lambda,s)
-\sigma_\lambda I_\lambda(\lambda,s) \nonumber \\
&+&\frac{\sigma_\lambda}{4\pi}\int_{\omega_i=0}^{4\pi} I_\lambda(\lambda,s, \omega_i)\Phi(\lambda,\omega,\omega_i)d\omega_i,
\ENA
where $\Phi$ is the phase function for scattering, $I_{\lambda,b}$ is the 
spectral intensity from a blackbody and 
$a_\lambda$ and $\sigma_\lambda$ are the spectral absorption and 
scattering coefficients, respectively.
For a medium in which only absorption is important, and where the absorption
coefficient is assumed to be constant for all wavelengths, 
the equation of radiative transfer reduces to
\EQ
\label{rad_trans2}
\frac{dI_\lambda(\lambda,s)}{ds}=-KI_\lambda(\lambda,s)+aI_{\lambda,b}(\lambda,s).
\EN
By neglecting
emission along the path the spectral intensity
of radiation after traveling a distance $s$ into a medium is then found
by integration of \Eq{rad_trans2} to be
\EQ
\label{intens}
I_\lambda(\lambda,s)=I_{\lambda}(\lambda,0)e^{-K s}.
\EN
Here $I_{\lambda}(\lambda,0)$ is the intensity at the beginning of the path,
the spectral intensity leaving a char particle, which is assumed to be
a graybody emitter.
For such radiation, the total intensity at a distance $s$ from the particle
is found by integrating over all wavelengths
\EQ
\label{intens2}
I(r)=\int_{\lambda=0}^\infty I_\lambda (\lambda,s)d\lambda
=\frac{\epsilon_p\sigma T_p^4}{\pi}e^{-Ks}.
\EN
Here, $\epsilon_p$ is the particle emissivity,
$\sigma$ is Stefan-Boltzmann constant and $T_p$ is the particle temperature. 
Particle scattering has been neglected
since for most relevant applications $E_a \gg E_s$.
Later in the paper,
the emission from each particle will be included 
through an integration over
spherical shells of increasing radius instead of through a direct
inclusion in the equation of radiative transfer. This does not result in any 
loss of generality and is done in order to simplify the calculations.

\begin{figure}[ht]\centering\includegraphics[width=1.0\textwidth]{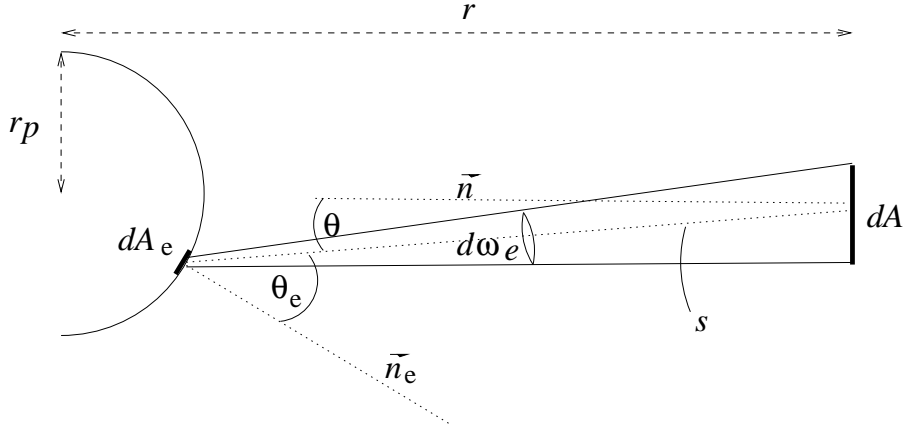}
\caption{Schematic of the variables used.\label{schematic}}
\end{figure}

The radiant energy $d^2Q$ per unit time in the small wavelength interval 
$d\lambda$ centered around $\lambda$ 
that is incident on a surface element $dA$ and 
originates from a surface element $dA_e$ on the surface of
a particle 
having a center a distance $r$ away from $dA$ is given by
\EQ
d^2Qd\lambda=I_\lambda(\lambda,r)d\omega_e\cos\theta_edA_ed\lambda
\EN 
where $d\omega_e$ is the solid angle subtended by $dA$ when viewed from $dA_e$
and is given by
\EQ
d\omega_e=\frac{\cos \theta dA}{s^2}.
\EN
Here $s$ is the distance between the differential elements 
$dA$ and $dA_e$ and $\theta_e$ and $\theta$ are
the angles between 
the straight line connecting $dA$ and $dA_e$ and
the normal to $dA_e$ ($\vec{n}_e$) and $dA$ ($\vec{n}$), respectively.
Shown in \Fig{schematic} is a schematic view of the variables.
Due to the curvature of the particle surface the distance between
$dA$ and $dA_e$ will generally be slightly different from $r$ and is 
denoted $s$. 

The total energy $dQ$ from the particle incident on $dA$ per time unit 
is found by integrating over all 
wavelengths and over the entire surface, $S_p$, of the particle;
\EQ
dQ=\int_{S_p} \int_{\lambda=0}^\infty I_\lambda(\lambda,s)
\frac{\cos \theta dA}{s^2}\cos\theta_e d\lambda dA_e
\EN
where $s$ will vary with $dA_e$ due to the curvature of the particle surface.
By assuming that $r_p<<r$, it follows that $s \rightarrow r$
and that $\theta$ becomes the angle between $\vec{n}$ and the line
connecting the center of the particle and $dA$. Now, by using
\Eq{intens2}, it can be found that
\EQ
dQ = \int_{\lambda=0}^\infty I_\lambda(\lambda,r) \frac{\pi r_p^2}{r^2} \cos\theta dA d\lambda
=\sigma T_p^4 \epsilon_p e^{-Kr}\left(\frac{r_p}{r}\right)^2 \cos\theta dA.
\EN
The flux at $dA$ due to radiation from the entire particle is now
\EQ
\label{Fbasic1}
q=\frac{dQ}{dA}= q_p e^{-Kr}\left(\frac{r_p}{r}\right)^2\cos\theta
\EN
where the radiative flux emitted from the surface of a particle is 
\EQ
\label{Fbasic}
q_p=\sigma T_p^4 \epsilon_p.
\EN
Assume now that $dA$ corresponds to the projected surface area 
of some particle {\bf p}$_c$ with radius $r_c$ and external surface are
$A_p=4\pi r_c^2$. The
total emission on {\bf p}$_c$ is then $q dA=q \pi r_c^2$, while $\theta=0$,
such that the mean flux $\mean{q}$ onto the surface of {\bf p}$_c$
due to a particle with radius $r_p$ placed a distance $r$ away from {\bf p}$_c$
is 
\EQ
\label{Fbasic2}
\mean{q}=\frac{\pi r_c^2 q}{A_p}=\frac{1}{4}q_p e^{-Kr}\left(\frac{r_p}{r}\right)^2.
\EN

\section{Solid-Solid radiation}

\subsection{Particle-wall radiation}
Let's now assume that we are in a spherical confinement with radius $R$.
The non-dimensional number 
$\tau=RK$
is the optical thickness.
In the case with negligible optical thickness ({\it i.e.} $\tau \rightarrow 0$)
the total radiative flux incident on the confinement wall due to the 
radiation from all particles is
\EQ
\label{fppwa0}
\lim_{\tau \rightarrow 0}q_{pp-w}
=\int_{r_p=0}^\infty\frac{\frac{4}{3}\pi R^3 n(r_p) \cdot 4\pi r_p^2q_p}{4\pi R^2}dr_p
=\frac{4}{3}\tau q_p
\EN
where the first term in the numerator yields the total number of particles
within the confinement and the second term give the emission from each of these
particles. Since negligible optical thickness is assumed,
it is clear that all radiation that is emitted from the particles will
eventually be incident on the wall, which explains why the denominator must
equal the surface area of the confinement wall in order to yield the
radiative {\it flux} on the wall. 

In the case of non-negligible optical thickness, the equation 
for the total radiative flux on the confinement walls 
is more complicated. 
Booth \cite{booth49} theoretically considered a cloud of radiating particles
in order to determine an effective emissivity that could be used to 
describe radiation from the particle cloud.
He showed that by assuming an absorption efficiency factor of unity,
the radiative emission incident on the walls surrounding the cloud, 
due to the enclosed particle cloud, is 
\EQ
\label{epwall}
q_{pp-w}=q_p \epsilon_{\rm eff}(\tau)
\EN
where
\EQ
\label{eq16}
\epsilon_{\rm eff}(\tau)=\left[1-\frac{1}{2\tau^2}+e^{-2\tau}
\left(
\frac{1}{\tau}+\frac{1}{2\tau^2}
\right)\right].
\EN
From this it is clear that the cloud of particles within the enclosure 
may be considered
as a single object with radius $R$, temperature $T_p$ and an effective emissivity
$\epsilon_{\rm eff}(\tau)$. 
For very small values of the optical thickness, it can be shown by 
Taylor expansion that
\EQ
\lim_{\tau\rightarrow 0}\epsilon_{\rm eff}(\tau)=\frac{4\tau}{3}.
\EN
such that in the case of vanishing $\tau$, \Eq{epwall} reduces to
\Eq{fppwa0}, as expected.

\section{Particle energy equation}
The energy conservation equation for a particle is given by
\EQ
\frac{dT_p}{dt}=\frac{1}{m_pc_{p,p}}(Q_{\rm con}+Q_{\rm rad}+Q_{\rm other})
\EN
where $T_p$ is the particle temperature, $m_p$ is the particle mass, $c_{p,p}$ is
the specific heat capacity of the particle and $Q_{\rm rad}$  
and $Q_{\rm con}$ represent the heating/cooling due to radiation and
convection and conduction, respectively, and $Q_{\rm other}$ represent
any other heating term that could be due to e.g. chemical reactions.
The effect of radiative absorption may be very important 
for the temperature evolution of a particle, but exactly how important 
the absorption is will depend on the position of the
particle within the particle cluster.

\subsection{Particle in the center of the enclosure}
At the end of \Sec{sec2}, $\bar{q}(r_p,r)$ was defined as the mean flux
at the surface of a particle due to the radiative emission from another 
particle with radius $r_p$ a distance $r$ away.
The total flux received by a particle in the center of the enclosure, $q_{pp-pc}$, 
is now found by integrating $\bar{q}(r_p,r)$ over all its surrounding particles.
This means by integration over all particle 
volumes $dV(r)$ and number densities $dn(r_p)$,
{\it i.e.}
\EQ
q_{pp-pc}=\int_{r_p=0}^\infty\int_{r=0}^R \bar{q}(r_p,r) dV(r)dn(r_p).
\EN
Since the volume of a spherical shell with thickness $dr$ and radius $r$ 
is $dV(r)=4\pi r^2 dr$, and since the particle number density of particles having
radii between $r_p$ and $r_p+dr_p$ is given by
$dn(r_p)=n(r_p)dr_p$, the above equation becomes
\EQ
\label{Fpppc}
q_{pp-pc}
=\int_{r_p=0}^\infty\int_{r=0}^R 4\pi r^2 \mean{q}(r_p,r) n(r_p)drdr_p
=q_{p}(1-e^{-\tau})
\EN
when \Eq{Fbasic2} is used for $\bar{q}(r_p,r)$ and all particles 
are assumed to behave alike.

The flux of radiation from the enclosure walls incident on the particle in the
center of the enclosure is 
\EQ
\label{Fwpc}
q_{w-pc}=(q_w+q_{w,r})e^{-\tau}
\EN
where the radiative flux emitted from a diffuse graybody wall is
\EQ
q_w=\epsilon_w\sigma T_w^4
\EN
and where the wall temperature and emissivity are given by $T_w$ and $\epsilon_w$,
respectively. The radiative flux reflected off the wall, $q_{w,r}$, 
is given by the 
product of the radiative flux received from the particles and the 
reflectivity of the wall, $\rho_w$, i.e.:
\EQ
q_{w,r}=\rho_w q_{pp-w},
\EN
where $q_{pp-w}$ is given by \Eq{epwall}.

The radiative cooling of the particle in the center of the particle cloud,
$Q_{\rm rad,centr}$,
is found by integrating the difference between the absorbed, $E_aq_{pc,rec}$, 
and the emitted, $q_{pc,em}$, radiative flux
over the particle surface of the particle in the center of the 
particle cloud. 
The radiative flux emitted from the particle is given by $q_{pc,em}=q_p$,
where $q_p$ is found from \Eq{Fbasic}, while the radiative flux
received by the particle in the center of the cloud is given by
the sum of the radiation received from the rest of the particle cloud and
the wall, i.e.
$q_{pc,rec}=q_{w-pc}+q_{pp-pc}$. 
Since the radiation in the
center of the spherical cloud is isotropic, such that the integration over
the particle surface can be replaced by the external particle surface area,
this yields
\EQ
\label{qrmcentr}
Q_{\rm rad,centr}=A_p(E_aq_{pc,rec}-q_{pc,em}),
\EN
where $A_p=4\pi r_p^2$ is the surface area of the particle.
By employing \Eq{Fpppc}, \Eq{Fwpc} and \Eq{qrmcentr}
the radiative cooling term of the particle in the center
of the particle cloud becomes
\EQ
\label{radterm}
Q_{\rm rad,centr}=
A_p\left(
  q_p\left[
    E_a(1+e^{-\tau}(\rho_w\epsilon_{\rm eff}-1))-1
  \right]+q_wE_ae^{-\tau} 
\right).
\EN

\subsection{Particle near the enclosure}
A particle very near the enclosure walls will receive the radiative flux from 
all the other particles on one side while on the other side it will
receive the flux from the wall. The mean flux received is therefore
$q_{pR}=\frac{1}{2}(q_{pp-w}+q_w+q_{w,r})$
which yields
\EQ
\label{radterm_R}
Q_{\rm rad,R}
=A_p\left[q_{pR}E_a-q_p \right]
=\frac{A_pE_a}{2}
\left[
q_p\epsilon_{\rm eff}(\tau)(1+\rho_w)+q_w
\right]
-A_pq_p.
\EN

\subsection{The ``mean'' particle}
In the following, a radiation term that on average will give the
correct net radiative outflow from the ``average'' particle in 
the cloud, is proposed.
The radiative term, $Q_{\rm rad,aver}$, 
is defined as the
net radiative flux from the entire particle cloud divided by the total 
number of particles in the cloud.

Since the gas is assumed not to take part in the radiative exchange, and
the container wall is assumed to be opaque, the only two radiatively 
active media are the particle cloud and the container wall. 
The net radiative heating of the wall, $E_{w,{\rm net}}$, equals the
radiation absorbed by the wall from the particles, minus the radiation
from the wall which is absorbed by the particles. Similarly the 
net radiative heating of the particles, $E_{p,{\rm net}}$, equals the
radiation absorbed by the particles from the wall, minus the radiation
from the particles which is absorbed by the wall. 
Based on this a radiative 
balance equation between the two media can be set up:
\EQ
\label{rad_equi}
E_{w,{\rm net}}=-E_{p,{\rm net}}.
\EN
Note that the above equation does {\it not} 
consider the energy balance of the system, it only states that the 
{\it net radiative} heating of the wall and the particles must sum to zero. 

Since all surfaces are assumed to be gray and diffuse
and since all particles are assumed to behave alike, the absorptivity of the particle cloud 
equals the effective emissivity found in \Eq{eq16}, $\epsilon_{\rm eff}(\tau)$, 
such that the total thermal emission from the wall incident on the particle
cloud is 
\EQ
\label{Ewpp}
E_{w-pp}=4\pi R^2q_w\epsilon_{\rm eff}(\tau).
\EN
The net radiative heating of the wall equals the radiative energy 
the wall absorbs from the 
particle cloud minus the radiative energy it emits as thermal radiation, i.e. 
\EQ
\label{ewnet}
E_{w,{\rm net}}=E_{pp-w}-E_{w-pp},
\EN
when $E_{pp-w}=4\pi R^2 q_{pp-w}\alpha_w$
and $\alpha_w=1-\rho_w$ is the absorptivity of the wall.
By using \Eq{epwall}, \Eq{Ewpp} and \Eq{ewnet}, it is found that 
the net radiative heating of the wall is
\EQ
\label{29}
E_{w,{\rm net}}
=4\pi R^2\epsilon_{\rm eff}(\tau)
\left(
\alpha_w q_p-q_w
\right).
\EN

In the beginning of this subsection 
the radiative cooling term of the {\it average} particle was defined as the
net radiative flux from the entire particle cloud divided by the total 
number of particles in the cloud.
This means that the integral of $Q_{\rm rad,aver}$ over all particles 
in the cloud must equal the negative of the net radiative 
heating of the particle cloud.
From this it is now clear that $Q_{\rm rad,aver}$ is found by
\EQ
\label{43}
E_{p,{\rm net}}=-\frac{4}{3}\pi R^3\int_{r_p=0}^\infty Q_{\rm rad,aver}(r_p)n(r_p)dr_p
\EN
when the cloud volume is given by $4\pi R^3/3$.
When using the relation 
\EQ
\label{44}
Q_{\rm rad,aver}=A_p q_{\rm rad,aver}=4\pi r_p^2q_{\rm rad,aver}, 
\EN
together with
\Eq{abs_coeff}, the integral in \Eq{43} is found to be
\EQ
\label{rad_int}
 \int_{r_p=0}^\infty Q_{\rm rad,aver}(r_p)n(r_p)dr_p
=4 q_{\rm rad,aver}\int_{r_p=0}^\infty n(r_p)\pi r_p^2 dr_p
=4q_{\rm rad,aver} K.
\EN 
Combining \Eq{rad_int} and \Eq{43} to eliminate the integral, and inserting
the resulting expression for $q_{\rm rad,aver}$ into \Eq{44} yields
\EQ
Q_{\rm rad,aver}
=-\frac{3A_pE_{p,{\rm net}}}{16K\pi R^3}.
\EN
Introducing \Eq{rad_equi} and \Eq{29} into the above results in the following expression
for the net radiative outflow from the ``average'' particle
\EQ
\label{qradaver}
Q_{\rm rad,aver}
=\frac{3\epsilon_{\rm eff}(\tau)A_p}{4\tau}\left(\alpha_w q_p-q_w\right)
\EN
since the optical depth of the enclosure is given by $\tau=KR$.
We propose that the use of this average radiative loss better approximates
the radiative loss of a particle in a particle cloud of particles compared
to previous methods neglecting the inter-particle radiation (\Eq{eq1}).
The proposed method is {\it not} applicable for CFD simulations of entire
combustors or reactors, where ordinary radiation models like e.g. the
discrete ordinates method or similar can be used.
Instead the proposed equation
is particularly useful when one is
not able to explicitly 
simulate the radiation from the full particle cloud but instead focus on
a single particle that is supposed to represent all the 
other particles. This is the case in the work of e.g. 
Qiao et al.~\cite{qiao12} and Mitchell et al.~\cite{mitchell07}.
The proposed radiative cooling term will also be applicable when 
Direct Numerical Simulations (DNS) are being used to simulate a very small 
sub domain of a real application\footnote{In a DNS all spatial and 
temporal scales of the fluid are fully resolved, hence the fundamental fluid 
equations can be solved without any modeling of the fluid equations. 
This yields very accurate and
reliable results, but it requires huge computational resources. With 
a DNS, even on the worlds largest computers, only small physical 
domains can therefore be considered. 

Note that for a typical DNS the embedded 
particles are assumed to be very small, and hence are not resolved. This means
that even though the fluid itself can be solved without any modeling, the 
fluid-particle coupling must be based on models, such as e.g. the Stokesian
drag law.
}. This is particularly so due to the small volumes realizable in
a DNS simulations, which requires a radiation model that does not need
access to the particles outside the small simulation volume.

\section{Importance of inter-particle radiation for some relevant configurations}
In the current section, 
a few examples of particle sizes and number densities as found in the
literature 
will be examined to investigate the importance of inter-particle 
radiation for some application. The cases studied have been kept simple 
in order to more easily isolate the
effect of particle number density, particle size and size of the 
enclosure on the particle cooling.
In  \Tab{table1}, particle data found in the literature is presented.
\begin{table}[h!]
\caption{Mean particle sizes and number densities from previous 
studies\cite{qiao12,park2012}. The listed extinction coefficients has been 
calculated from \Eq{ext_coeff}.}
\label{table1}
\begin{tabular}{ccccc}
\hline\noalign{\smallskip}
Case & Reference &$n_p$ [m$^{-3}$] & $r_p$ [m]& $K$ [m$^{-1}$]\\
\noalign{\smallskip}\hline\noalign{\smallskip}
A&Qiao et al. (2012) \cite{qiao12}& $1\times 10^{9}$ & $5\times 10^{-5}$ & 8 \\
B&Park et al. (2012) \cite{park2012}& $5\times 10^{9}$ & $1.25\times 10^{-5}$ & 2.5 \\
C&Park et al. (2012) \cite{park2012}& $4\times 10^{8}$ & $1.25\times 10^{-5}$ & 0.2 \\
\noalign{\smallskip}\hline
\end{tabular}
\end{table}
Case A is from a coal gasification reactor, while the
data of \cite{park2012} are from two different locations in a pulverized 
coal furnace: the lower part of the furnace close to the burners (Case B)
and the upper part of the furnace, downstream of the burners, where
temperatures are relatively low (Case C). 

\begin{figure}[ht]\centering\includegraphics[width=1.0\textwidth]{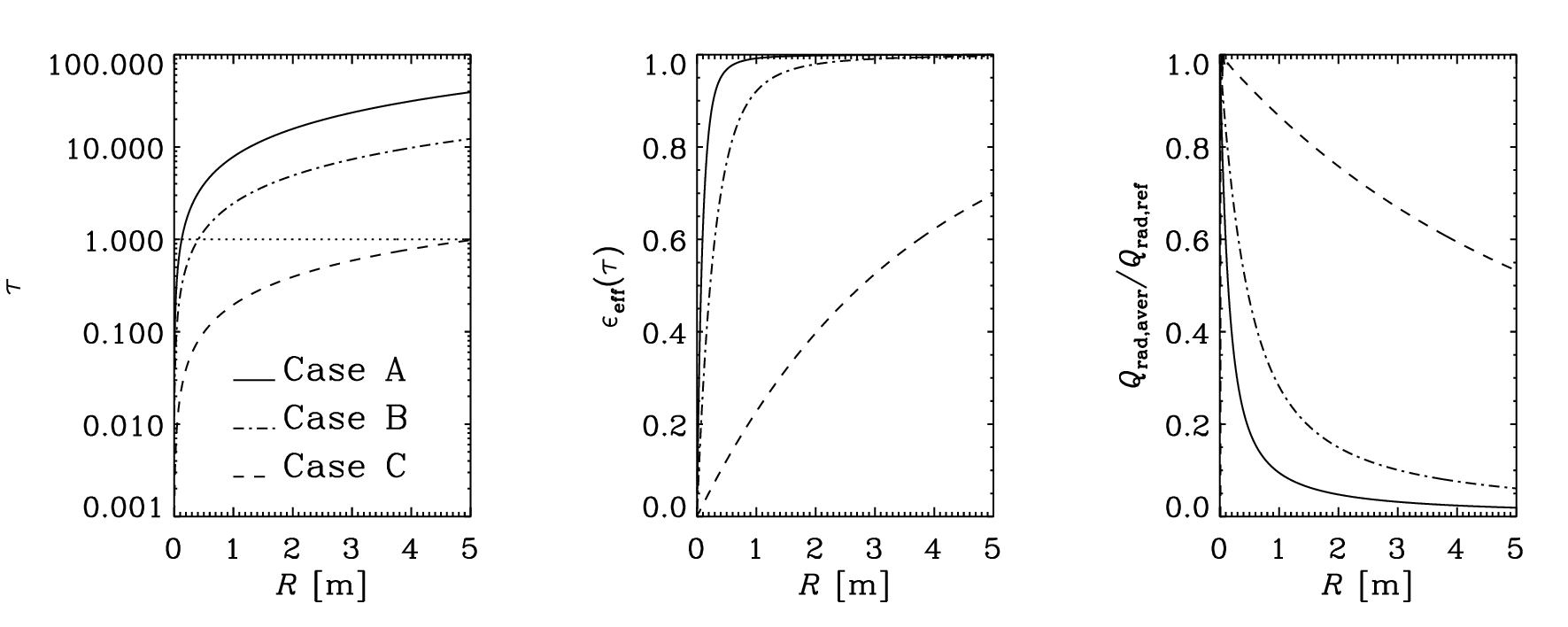}
\caption{Optical depth, $\tau$, (left),
effective emissivity of the particle cloud, $\epsilon_{\rm eff}(\tau)$, 
(middle) and 
normalized net radiative cooling of the ``average'' particle, 
$Q_{\rm rad,aver}/Q_{\rm rad,ref}$ (right).
All results are for a single particle size, {\it i.e.} $\sigma_p=0$, where
$r_p$ is given in \Tab{table1}.
Particle and wall temperatures have been set to 1200~K and 500~K, respectively.
\label{Qrad_fig}}
\end{figure}
In the left panel of \Fig{Qrad_fig} the optical thickness is 
plotted as a function of the enclosure radius $R$ for all three cases
listed in \Tab{table1}. The inter-particle radiation is important when
$\tau\gtrsim 1$, 
which is marked with a horizontal dotted line in the figure,
so for case C, inter-particle radiation starts to have a significant effect
for $R\gtrsim 5$~m. For case A and B
inter-particle radiation becomes important when the radius 
of the domain
exceeds about 10~cm and 30~cm, respectively. 

In the central panel, the absorption efficiency factor
of the particle cloud is shown as 
a function of enclosure radius.
For case A and B the emissivity is seen to approach unity for enclosure radii 
of 1~m and 3~m, respectively. This means that for radii above this the
particle cloud essentially behaves as a solid body with 
temperature $T_p$ and radius $R$. The same is not true for case C, which 
for all radii considered behaves like a cloud of diluted radiating 
particles.

In the right panel 
$Q_{\rm rad,aver}$ normalized by a
reference cooling term $Q_{\rm rad,ref}$ is shown. 
Here the reference cooling term is
obtained by neglecting particle-particle radiation, {\it i.e.}
\EQ
\label{qref}
Q_{\rm rad,ref}=A_p(q_p-E_aq_w).
\EN
From this it is clear that for large and/or dense particle clouds, the
average radiative cooling for the particles 
is much weaker than when inter-particle radiation is neglected. For example,
for case~A
with an enclosure radius of 2~m the reference cooling term  
is a factor 20 stronger than the
cooling term for the average particle.

\begin{figure}[ht]\centering\includegraphics[width=1.0\textwidth]{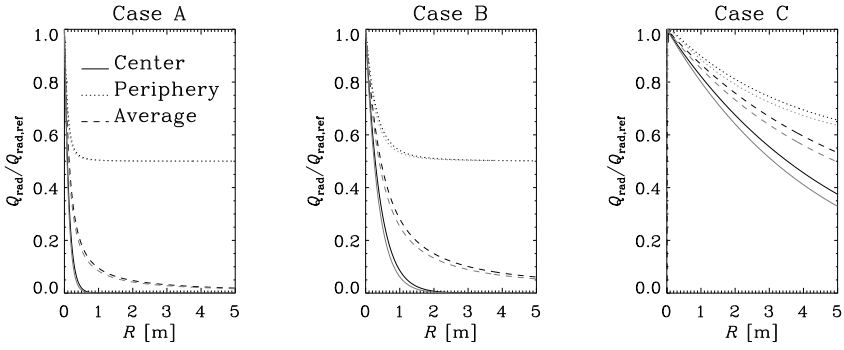}
\caption{Here $Q_{\rm rad}/Q_{\rm rad,ref}$ is plotted as a function of 
the radius of the enclosure for particles positioned in the center and on 
the periphery of the domain together with the value for the average of all
the particles. The three different panes represent the three different
cases listed in \Tab{table1}. 
The grey lines correspond to a particle distribution width 
$\sigma_p=\mean{r}_p\gamma_p$
where $\gamma_p=0.2$, while for the black lines $\gamma_p=0$.
Particle and wall temperatures have been set to 1200~K and 500~K, respectively.
\label{Qrad3_fig}}
\end{figure}

In \Fig{Qrad3_fig} the radiative cooling of a particle normalized by
the reference cooling given by \Eq{qref} is plotted as a function of enclosure
radius for different particle positions within the enclosure. 
The different position are 1) the
center of the domain, given by \Eq{radterm} (solid line), 2) the periphery, 
given by \Eq{radterm_R} (dotted line) and 3) the 
position of the average particle, given by \Eq{qradaver}, (dashed line). 
It is clearly seen that the cooling
is largest at the periphery, but that the difference is much less for case C 
where the particle number density is much smaller.
Furthermore it is interesting to note that the average cooling approaches zero
even for an enclosure radius of 5~m for case A and that the central particles
of the same case experience near zero cooling even for enclosure radii less than 
a meter. 

The grey lines in \Fig{Qrad3_fig} represent a distribution 
width of $\sigma_p=0.2 r_p$
while the black lines represent $\sigma_p=0$. As can be seen, 
the radiation term is not very sensitive to the width of the
particle size distribution even for a width as wide as 20\% of the mean 
particle radius. The effect of the broader particle size distribution is
largest for small optical depths, as in Case C, but
even here it is rather small.

Simulations of the gasification process presented in a paper by 
Qiao et al. (2012)~\cite{qiao12} has been performed
in order to emphasize the importance of  
including inter-particle radiation for dense clouds of particles.
The numerical code used to perform the simulations was comparable to the
code used in the above mentioned paper. 
Tests were done both with the same
radiative cooling term as used by Qiao et al. (\Eq{qref}), which 
neglects inter-particle radiation, 
and with the particle cooling term as proposed in this work 
(\Eq{qradaver}), which includes inter-particle radiation. Compared to when
inter-particle radiation is included, as given by \Eq{qradaver},
the time required to reach full conversion of the char is  47\% longer when
inter-particle radiation is neglected (\Eq{qref}).

Analytical expressions for geometries of the confinement walls
other than the spherical geometry 
considered in this work do not exist. 
It can be shown\cite{Field67}, however,
that other geometries like cylinders or cubes give trends
for the heat transfer that are similar to what is found for spherical geometries.
In particular it can be shown by numerical integration~\cite{Field67,erkku} 
that for cubes and
cylinders having aspect ratios near unity, the expressions developed for
spherical geometries give comparable results for the
net heat transfer to the enclosure walls.
It is therefore assumed
to be a good approximation to use the expressions developed here also
for real applications such as furnaces.

\section{Conclusion}
The particle cooling due to radiation has been investigated in particle 
clusters of variable size. When neglecting the effect of scattering and 
assuming all particles to behave alike it is shown that the
radiative particle cooling is very sensitive to where the particle is 
positioned within the particle cluster. 
Broadening the particle size distribution is found to just have a minor
impact on the results presented.

Instead of the traditional particle cooling term often used for
single particle simulations of particles in a cluster of particles (\Eq{qref})
a new particle cooling term is proposed (\Eq{qradaver})
where the particle cooling is defined as the average particle cooling of 
all the particles. In contrast to \Eq{qref}, the new
particle cooling term does include inter-particle radiation, which is found
to be very important for the applications studied.

We claim that, compared to previous methods that neglect the 
inter-particle radiation, the use of the proposed radiative 
cooling term better approximates
the radiative loss of a particle in a cloud of particles.
The proposed method is applicable for simulations of {\it small sub-volumes} of
gasifiers, pulverized coal combustors or any system where hot particle 
clouds exists. It is particularly useful when one is
not interested in simulating the radiation from the full particle cloud 
but instead want focus on
a single particle that represent all the 
other particles in the sub volume. Examples of such simulations are found in  
Qiao et al.~\cite{qiao12} and Mitchell et al.~\cite{mitchell07}.

\section*{Acknowledgements}
This work forms part of the CAMPS project supported by the Research
Council of Norway (215707). The work has additionally been produced
with support from the BIGCCS Centre, performed under the Norwegian
research program Centres for Environment-friendly Energy Research
(FME). The authors acknowledge the following partners for their
contributions: Aker Solutions, ConocoPhillips, Gassco, Shell, Statoil,
TOTAL, GDF SUEZ and the Research Council of Norway (193816/S60).

{}

\end{document}